\documentclass[twocolumn,aps,prc,showpacs,superscriptaddress,floatfix,nofootinbib]{revtex4-2}

\usepackage{newtxtext,newtxmath}
\usepackage{booktabs,siunitx}
\usepackage{dcolumn}
\usepackage{bm}
\usepackage{float}
\usepackage{ulem}
\usepackage{graphicx}
\usepackage{tabularx}
\usepackage{amsmath}
\usepackage{xcolor}
\usepackage{multirow}
\usepackage[colorlinks,citecolor=blue,urlcolor=blue,linkcolor=blue]{hyperref}

\newcommand{\pT}{p_{\rm T}}
\newcommand{\rhopearson}{\rho(v_{n}^{2}, \delta \pT)}
\newcommand{\rhotwo}{\rho(v_{2}^{2}, \delta \pT)}
\newcommand{\rhothree}{\rho(v_{3}^{2}, \delta \pT)}
\newcommand{\deltapT}{\langle(\delta \pT)^{2}\rangle}

\newcommand {\vnt} {v_{n}\{2\}}
\newcommand{\snn}{\sqrt{s_{\rm NN}}}
\newcommand{\Pb}{$^{208}$Pb}
\newcommand{\Ca}{$^{48}$Ca}
\newcommand{\PbPb}{$^{208}$Pb+$^{208}$Pb}
\newcommand{\CaCa}{$^{48}$Ca+$^{48}$Ca}

\newcommand {\RuRu} {$^{96}_{44}$Ru+$^{96}_{44}$Ru}
\newcommand {\ZrZr} {$^{96}_{40}$Zr+$^{96}_{40}$Zr}
\newcommand {\mean}[1] {\langle #1\rangle}             
\newcommand {\meanpt} {\mean{[\pT]}}  
\newcommand{\trento}{T\raisebox{-0.5ex}{R}ENTo}
\newcommand{\iebe} {{\tt iEBE-VISHNU}}

\begin{document}


\title{Impact of Geometric Inflation on Nucleon Size Sensitivity in Relativistic Heavy-Ion Collisions}

\author{Jian-fei Wang}
\affiliation{Department of Modern Physics, University of Science and Technology of China, Anhui 230026, China}
\affiliation{School of Science, Huzhou University, Huzhou, Zhejiang 313000, China}

\author{Hao-jie Xu}
\email{haojiexu@zjhu.edu.cn}
\affiliation{School of Science, Huzhou University, Huzhou, Zhejiang 313000, China}
\affiliation{Strong-Coupling Physics International Research Laboratory (SPiRL), Huzhou University, Huzhou, Zhejiang 313000, China.}

\begin{abstract}
	The intrinsic transverse size of nucleons, parameterized by a Gaussian width~$w$, is a critical yet uncertain input in the initial-state modeling of relativistic heavy-ion collisions. Using a finite~$w$ in standard initial geometry models introduces an unintentional ``geometric inflation'' that alters the initial nuclear density profile. In this study, we implement a self-consistent density correction to eliminate this artifact and investigate its impact on final-state observables. Through hybrid (viscous hydrodynamics + hadronic transport) simulations of $^{208}$Pb+$^{208}$Pb collisions at the LHC, we demonstrate that removing geometric inflation significantly modifies the sensitivity of observables to the nucleon width~$w$. While elliptic flow and mean transverse momentum ($\langle [p_{\rm T}]\rangle$) become less sensitive to variations in~$w$, the Pearson correlation coefficient~$\rho(v_{n}^{2}, \delta \pT)$, $[p_{\rm T}]$ fluctuations, and triangular flow exhibit enhanced sensitivity to fluctuations in nucleon positions. Our results indicate that uncorrected geometric inflation can bias the extraction of nucleon structure and quark-gluon plasma properties. This underscores the necessity of a self-consistent initial-state geometry for reliable Bayesian inference in heavy-ion collisions.
\end{abstract}

\maketitle

\section{Introduction}

Determining the intrinsic spatial scales of the nucleon is a central objective in nuclear and particle physics~\cite{Pohl:2010zza,Tiesinga:2021myr,ParticleDataGroup:2024cfk}. The proton charge radius has been measured with high precision through electron scattering and muonic hydrogen spectroscopy, providing a well-determined reference for the electromagnetic structure~\cite{A1:2010nsl,Antognini:2013txn}. However, the spatial scales governing strong-interaction dynamics—such as the gluonic radius or the mass radius—remain poorly constrained. These scales are of particular interest for high-energy scattering processes. For instance, diffractive $J/\psi$ photoproduction probes a gluonic radius of approximately $0.50$~fm~\cite{Caldwell:2010zza}, while near-threshold vector-meson photoproduction suggests a mass radius around $0.55$~fm~\cite{Kharzeev:2021qkd}. An improved understanding of these scales is essential for interpreting observables in heavy-ion collisions (HIC) and at the future Electron–Ion Collider (EIC).

In the context of heavy-ion collisions, the effective transverse size of the constituent nucleons is commonly parameterized by a Gaussian width, $w$~\cite{Moreland:2014oya,Loizides:2017ack}. Relativistic hydrodynamic simulations of the Quark--Gluon Plasma (QGP) have traditionally utilized widths of $w \approx 0.4$\,fm, a value consistent with the aforementioned gluonic and mass radii. However, recent Bayesian analyses, which treat initial state parameters and medium properties simultaneously, have inferred significantly larger values, with $w \approx 0.9$--$1.1$\,fm~\cite{Moreland:2018gsh,Bernhard:2019bmu,Nijs:2020roc,JETSCAPE:2020mzn}. This creates a significant discrepancy between phenomenological extractions and expectations from fundamental QCD processes. Resolving this tension is critical, as the nucleon width dictates the granularity of initial energy-density profiles, thereby influencing the extraction of QGP transport coefficients, such as the specific shear viscosity~\cite{Song:2017wtw}.

To resolve this ambiguity, recent studies have identified a highly discriminating observable: the Pearson correlation coefficient, $\rhopearson$, between anisotropic flow harmonics $v_n$ and the event-mean transverse momentum $[\pT]$. Unlike standard flow observables, this correlation exhibits a unique sensitivity to the nucleon width while remaining largely insensitive to medium viscosity~\cite{Giacalone:2021clp,Giacalone:2020dln}. Hydrodynamic calculations demonstrate that $\rhopearson$ changes substantially—even flipping sign—as $w$ varies from $0.4$\,fm to $1.2$\,fm. Comparisons with LHC data from the ALICE and ATLAS collaborations favor smaller widths ($w \approx 0.4$--$0.5$\,fm)~\cite{ALICE:2021gxt,ATLAS:2019pvn} and strongly disfavor the larger values suggested by Bayesian analyses ($w \gtrsim 0.8$\,fm)~\cite{Moreland:2018gsh,Bernhard:2019bmu,Nijs:2020roc,JETSCAPE:2020mzn}. Consequently, $\rhopearson$ has emerged as a promising, viscosity-independent probe of the initial state geometry, capable of breaking parameter degeneracies and bridging HIC measurements with deep inelastic scattering data~\cite{Giacalone:2021clp}.

However, the interpretation of these geometrical effects is complicated by a systematic artifact known as ``geometric inflation''~\cite{Xu:2026vnl}. This effect arises when finite nucleon widths are applied to discrete nucleon positions sampled from a standard Woods-Saxon distribution, unintentionally inflating the global nuclear radius and diffuseness. This necessitates a revision of the theoretical baseline for anisotropic flow harmonics, $v_n$. The impact of geometric inflation is not limited to global cross-sections~\cite{Xu:2026vnl}; it fundamentally biases the initial spatial eccentricities. In this study, we extend the self-consistent density correction framework to relativistic hydrodynamic simulations to rigorously quantify the effects of geometric inflation on both $v_n$ and the correlation $\rhopearson$. Our goal is to determine whether the reported sensitivity of these observables to nucleon size is a genuine physical signal or a consequence of unintentional geometric rescaling. Establishing this distinction is indispensable for the next generation of Bayesian analyses, ensuring that the extraction of QGP transport properties—such as shear and bulk viscosities—is not compromised by systematic biases in initial state modeling.

The paper is organized as follows. In Sec.~\ref{sec:model}, we introduce the self-consistent nuclear density for heavy-ion collisions and the Pearson coefficient used in this study. The hydrodynamic results and their initial predictors are discussed in Sec.~\ref{sec:discussion}, and a summary is provided in Sec.~\ref{sec:summary}.

\section{Model and Setup}
\label{sec:model}

\subsection{Self-consistent Initial State Geometry}

Geometric inflation arises from the inconsistency between point-like sampling of nucleon positions and their subsequent convolution with finite-sized profiles. In a standard Monte Carlo Glauber approach, nucleon positions are sampled from a Woods-Saxon distribution:
\begin{equation}
    \rho^{\mathrm{WS}}(r) = \frac{\rho_0}{1+\exp[(r-R)/a]},\label{W-S}
\end{equation}
with $R=6.62$\,fm and $a=0.546$\,fm for the $^{208}\mathrm{Pb}$ nucleus~\cite{Loizides:2017ack,Moreland:2014oya}. When these positions are convolved with a Gaussian profile of width $w$, the resulting effective density $\rho(r;w)$ deviates from $\rho^{\mathrm{WS}}$, leading to an unintentional inflation of the root-mean-square radius and a broadening of the surface diffuseness.

To eliminate this artifact, we adopt the self-consistent framework proposed in Ref.~\cite{Xu:2026vnl,Klos:2007is,Salcedo:1987md,Oset:1989ey,Xu:2021vpn}, which preserves the target nuclear density by modifying the sampling distribution $f(\vec{\xi})$ via an inverse Weierstrass transform~\cite{Yang:2021gwa,Xu:2026vnl}:
\begin{equation}
    \rho(\vec{r};w) = \frac{1}{(2\pi w^2)^{3/2}} \int d^3\xi \, f(\vec{\xi}) \,
    \exp\!\left(-\frac{(\vec{r} - \vec{\xi})^2}{2w^2}\right).
    \label{eq:weierstrass}
\end{equation}
By requiring that the first three radial moments of $\rho(\vec{r};w)$ match those of the physical $\rho^{\mathrm{WS}}(\vec{r})$ distribution, we determine the corrected sampling parameters $(\tilde{R}, \tilde{a})$ of the Woods-Saxon approximated $f(\vec{\xi})$ for a given nucleon width $w$~\cite{Xu:2026vnl}. In this study, we focus on a representative large width of $w=0.92$\,fm, motivated by Bayesian extractions~\cite{JETSCAPE:2020mzn,Nijs:2020roc,Bernhard:2019bmu}.
To maintain the physical nuclear geometry at this width, we utilize the corrected sampling parameters $\tilde{R} = 6.72$\,fm and $\tilde{a} = 0.24$\,fm. Additional sampling parameters for different $w$ values are listed in Tab.~\ref{tab:paras}. This procedure ensures that the global nuclear geometry remains invariant under smearing, effectively decoupling genuine nucleon-size effects from unintentional geometric rescaling. For a large width of $w=0.92$ fm, the correction results in a point-nucleon distribution that is more compact ($\tilde{a}=0.24$ fm) than the original Woods-Saxon ($a=0.546$ fm), compensating for the subsequent smearing.

\subsection{Hydrodynamic Framework and Observables}

We simulate relativistic \PbPb\ collisions at $\snn = 5.02$\,TeV using the \iebe\ hybrid model~\cite{Song:2010mg,Shen:2014vra,Shen:2020mgh}. The framework integrates \trento~\cite{Moreland:2014oya} for the initial energy deposition, $(2+1)$D viscous hydrodynamics (\texttt{VISH2+1})~\cite{Song:2007ux} for the QGP expansion, and \texttt{UrQMD}~\cite{Bass:1998ca,Bleicher:1999xi} for the late-stage hadronic evolution. The physical parameters of the medium, including viscosities and switching temperature, are set to the maximum a posteriori (MAP) values obtained from the Bayesian calibration in Ref.~\cite{Moreland:2018gsh}
~\footnote{In this Bayesian analysis, the sub-nucleon fluctuation mode with a consistent number $m=6$ is calibrated. Incorporating such fluctuations can modify both the nucleon thickness function and the effective partonic cross section, which are otherwise corrected in the default \trento\  model. Here we fix the sub-nucleon fluctuations to their MAP values and leave a more detailed study of these fluctuations for future work.}.

Event centrality is determined by the final-state charged-particle multiplicity within the experimental acceptance. To match experimental protocols, we calculate flow and correlation observables for charged hadrons within $0.2 < \pT < 3.0$\,GeV.

To suppress non-flow contributions and avoid trivial self-correlations in the calculation of $\rhopearson$, we employ the 3-subevent method. The detector acceptance is divided into three distinct regions in pseudorapidity: subevent $A$ ($\eta < -0.6$), subevent $B$ ($-0.6 < \eta < 0.6$), and subevent $C$ ($\eta > 0.6$). The Pearson correlation coefficient is then defined as~\cite{Bozek:2016yoj, Schenke:2020uqq, Giacalone:2020dln}:
\begin{equation}
    \rhopearson = \frac{\text{Cov}(v_n^2, [\pT])}{\sqrt{\text{Var}(v_n^2)_{\mathrm{dyn}} \text{Var}([\pT])_{\mathrm{dyn}}}}.
\end{equation}
The covariance $\text{Cov}(v_n^2, [\pT])$ is calculated using $v_n^2$ from particles in subevents $A$ and $C$, and $[\pT]$ fluctuations from particles in subevent $B$. The flow harmonics are calculated with the Q-cumulant method, \(c_{n} = v_{n}\{2\}^{2}= \langle\langle 2\rangle\rangle_{n,-n}\), using the correlator
\[
\langle\langle m \rangle\rangle_{n_1,\ldots,n_m} \equiv \left\langle e^{i(n_1\varphi_1 + \cdots + n_m\varphi_m)} \right\rangle.
\]
The quantities are averaged over the particles of interest (POIs) within an event by the inner $\langle \cdots \rangle$, and over all events by the outer $\langle \cdots \rangle$~\cite{Bilandzic:2010jr,Bilandzic:2013kga}. The $[\pT]$ fluctuation is obtained with~\cite{STAR:2005vxr,Giacalone:2020lbm} 
\begin{equation}
    \deltapT= \left\langle\frac{Q_{1}^{2} - Q_2}{N(N-1)} \right\rangle - \left\langle \frac{Q_1}{N} \right\rangle^{2},
\end{equation}
where $Q_{n} = \sum_{i=1}^{N} p_{{\rm T},i}$ for a given event.

\begin{table}[b]
\caption{The corrected Woods-Saxon parameters for sampling nucleon centers ensure that the global nuclear geometry remains invariant under Gaussian smearing with parameter $w$.} \label{tab:paras}
\setlength{\tabcolsep}{8pt}
\renewcommand{\arraystretch}{1.2} 
\begin{tabular}{c c c c c}   
\toprule
\multirow{2}{*}{$w$ (fm)} & \multicolumn{2}{c}{\Pb} & \multicolumn{2}{c}{\Ca} \\
\cline{2-5}   
 & $\tilde{R}$ (fm) & $\tilde{a}$ (fm)& $\tilde{R}$ (fm)& $\tilde{a}$ (fm)\\
\midrule
0.4 & 6.64    & 0.504     & 3.73      & 0.487  \\
0.5 & 6.65    & 0.479     & 3.74      & 0.464  \\
0.6 & 6.66    & 0.445     & 3.74      & 0.435  \\
0.7 & 6.67    & 0.402     & 3.75      & 0.396  \\
0.8 & 6.70    & 0.344     & 3.77      & 0.345  \\
0.9 & 6.71    & 0.262     & 3.80      & 0.271  \\
1.0 & 6.74    & 0.109    & 3.84     & 0.138  \\
\bottomrule
\end{tabular}
\end{table}

\begin{figure*}[!htb] 
\centering
\includegraphics[width=0.45\textwidth]{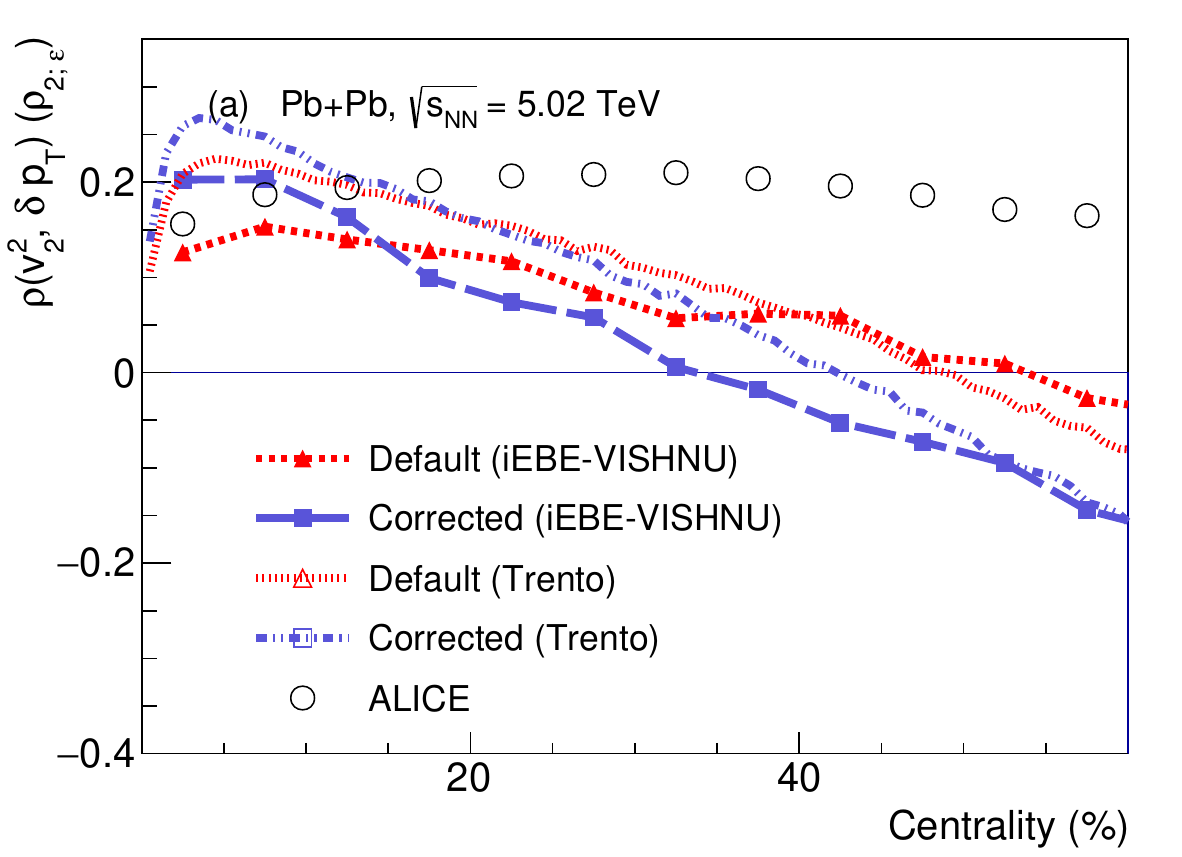}
\includegraphics[width=0.45\textwidth]{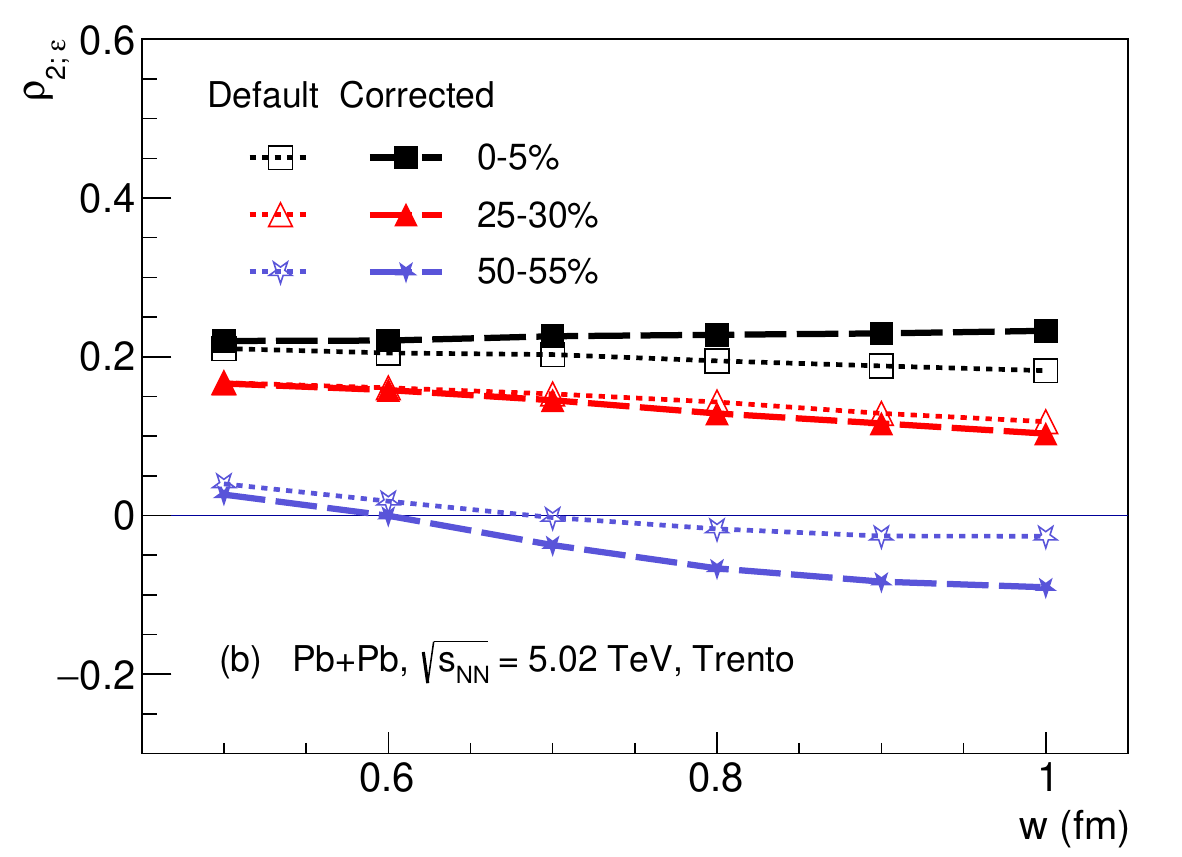}
\includegraphics[width=0.45\textwidth]{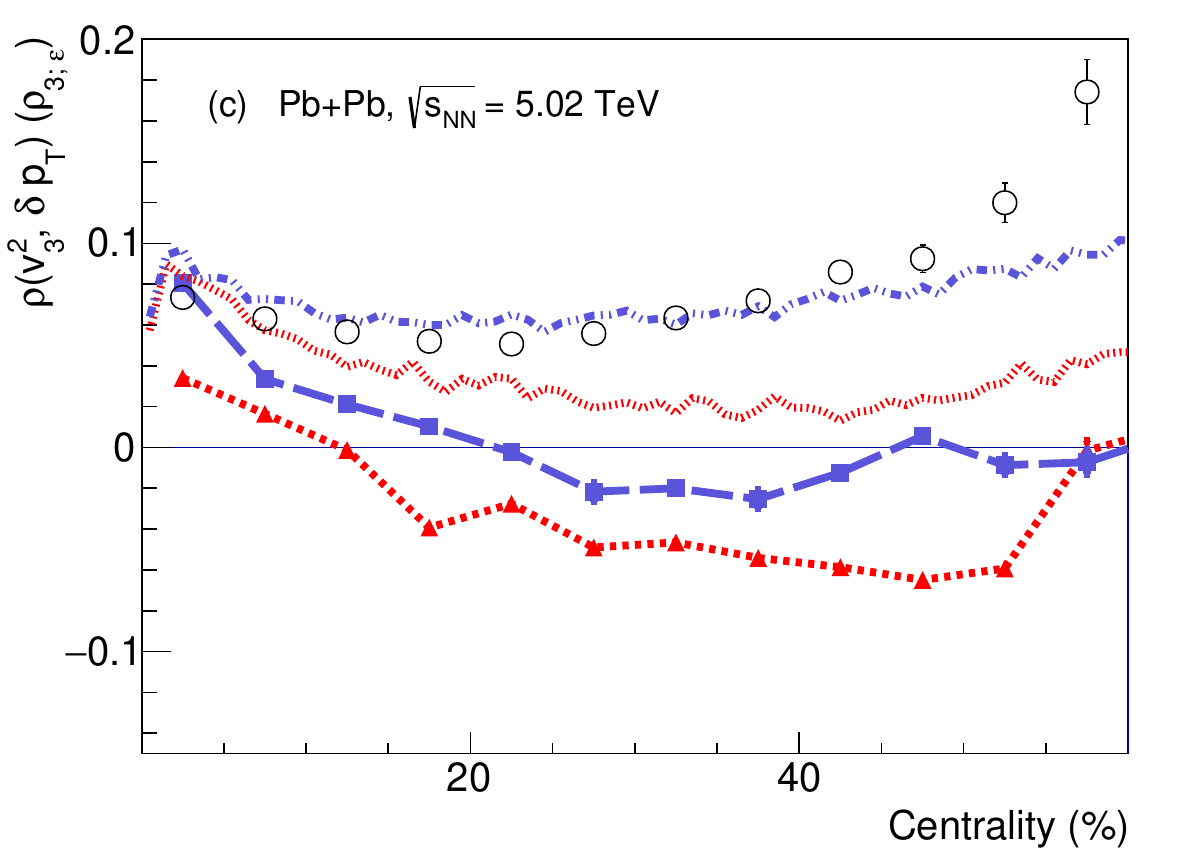}
\includegraphics[width=0.45\textwidth]{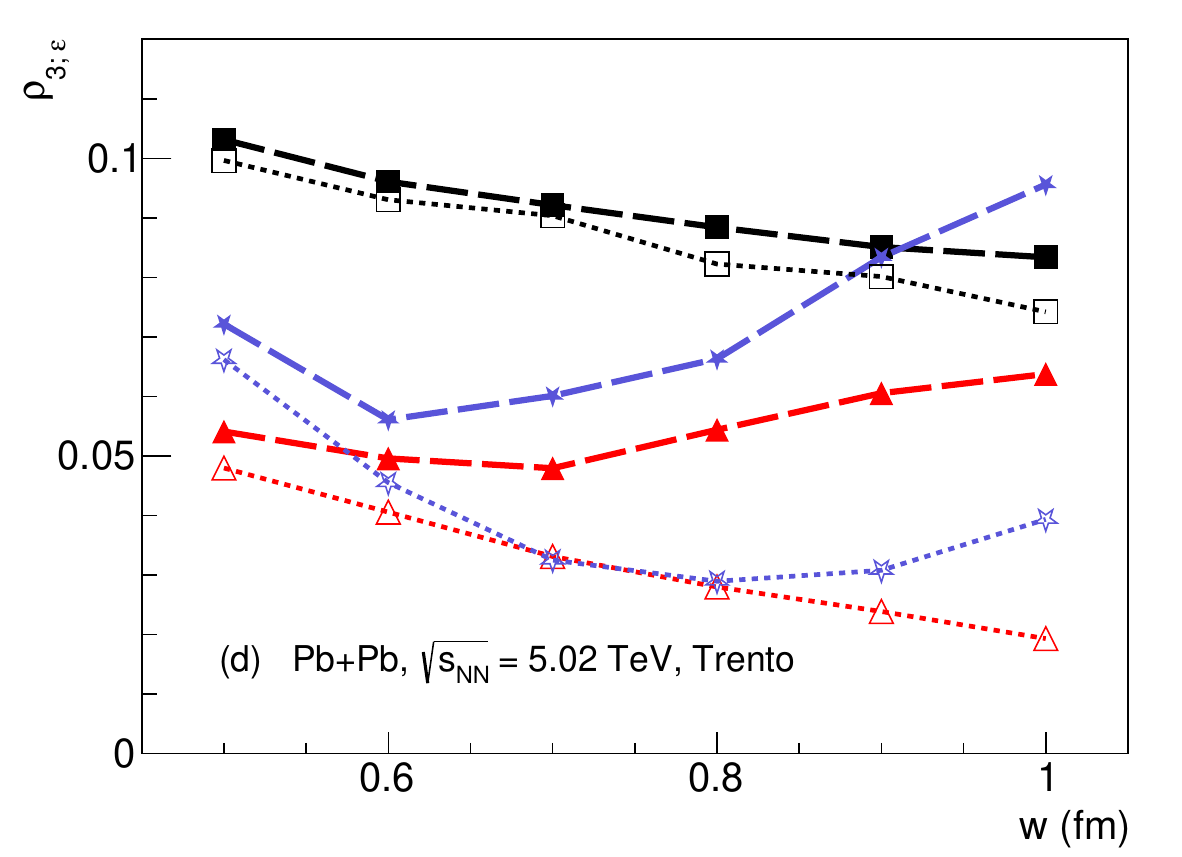}
\caption{(Color online) (Left panels) Centrality dependence of $\rhotwo$ and $\rhothree$ with and without geometric inflation corrections in \PbPb\ collisions at $\snn=5.02 \ \mathrm{TeV}$ from \iebe\ simulations. The right panels show the nucleon-size dependence of the corresponding initial predictors ($\rho_{2;\varepsilon}$ and $\rho_{3;\varepsilon}$) from \trento\ simulations, which capture the relative differences between the two scenarios. Experimental data are from Ref.~\cite{ALICE:2021gxt}.} \label{fig:rhohydro}
\end{figure*}

\begin{figure*}[thb]
\centering
\includegraphics[width=0.45\textwidth]{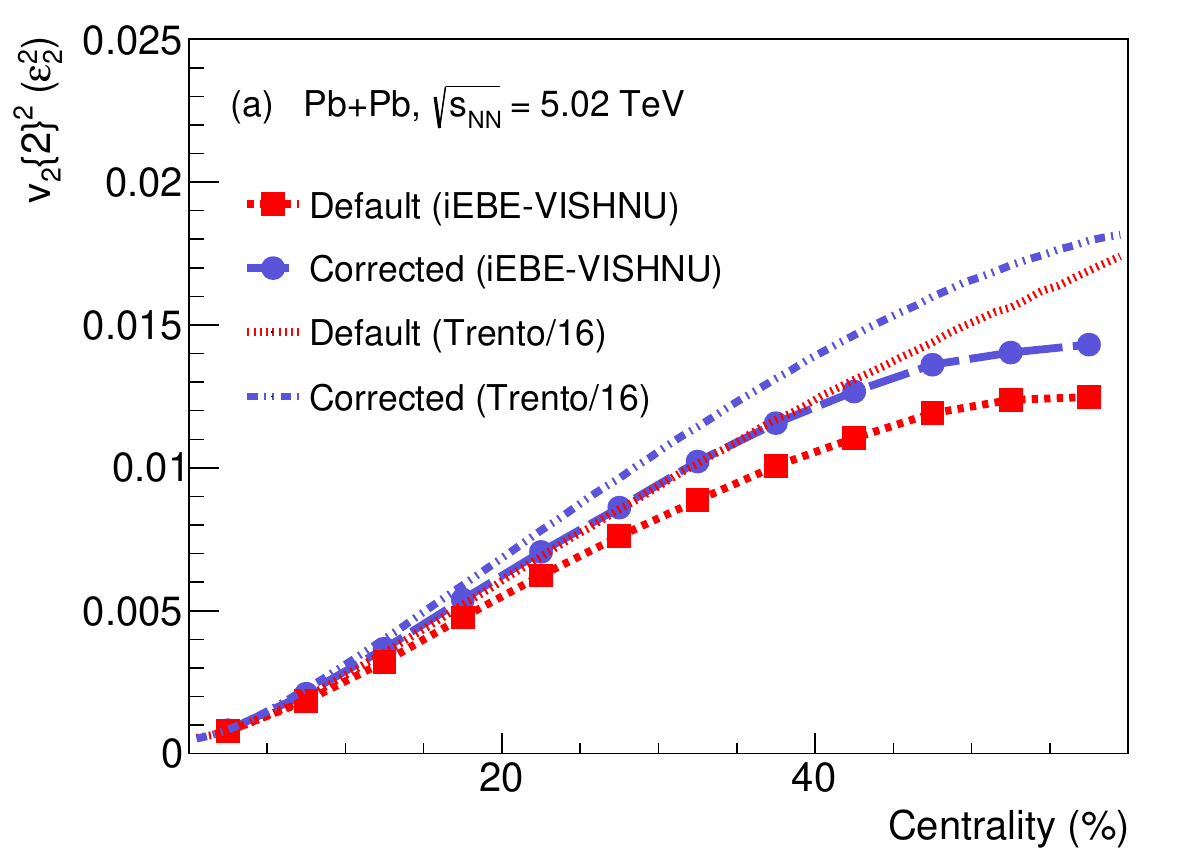}
\includegraphics[width=0.45\textwidth]{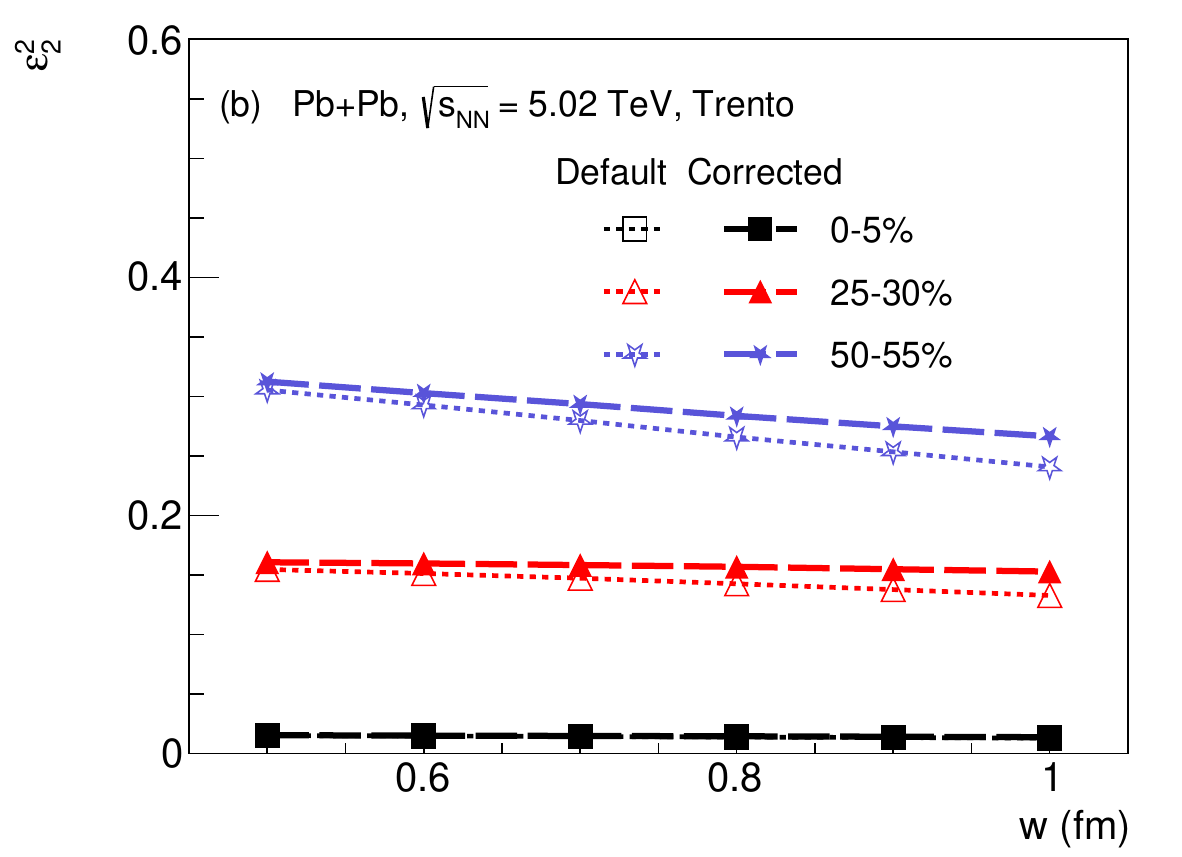}
\includegraphics[width=0.45\textwidth]{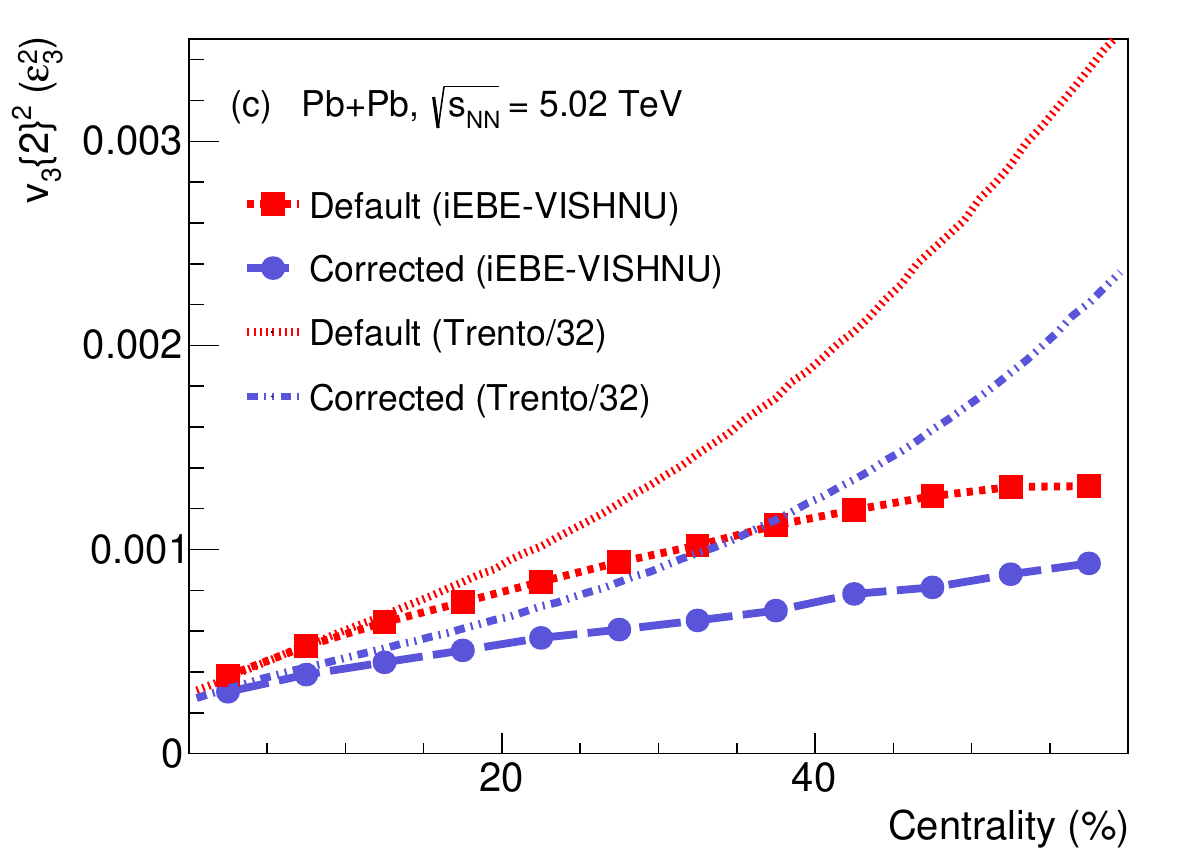}
\includegraphics[width=0.45\textwidth]{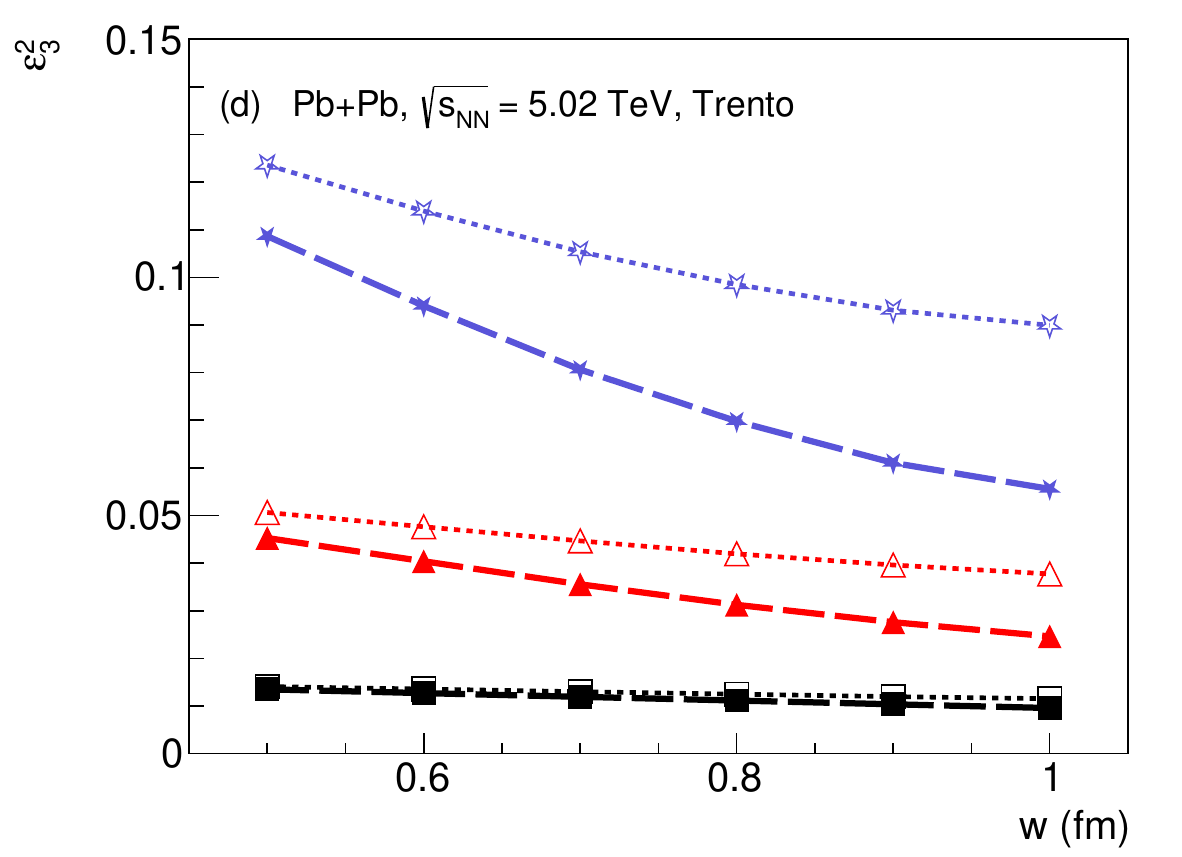}
\includegraphics[width=0.45\textwidth]{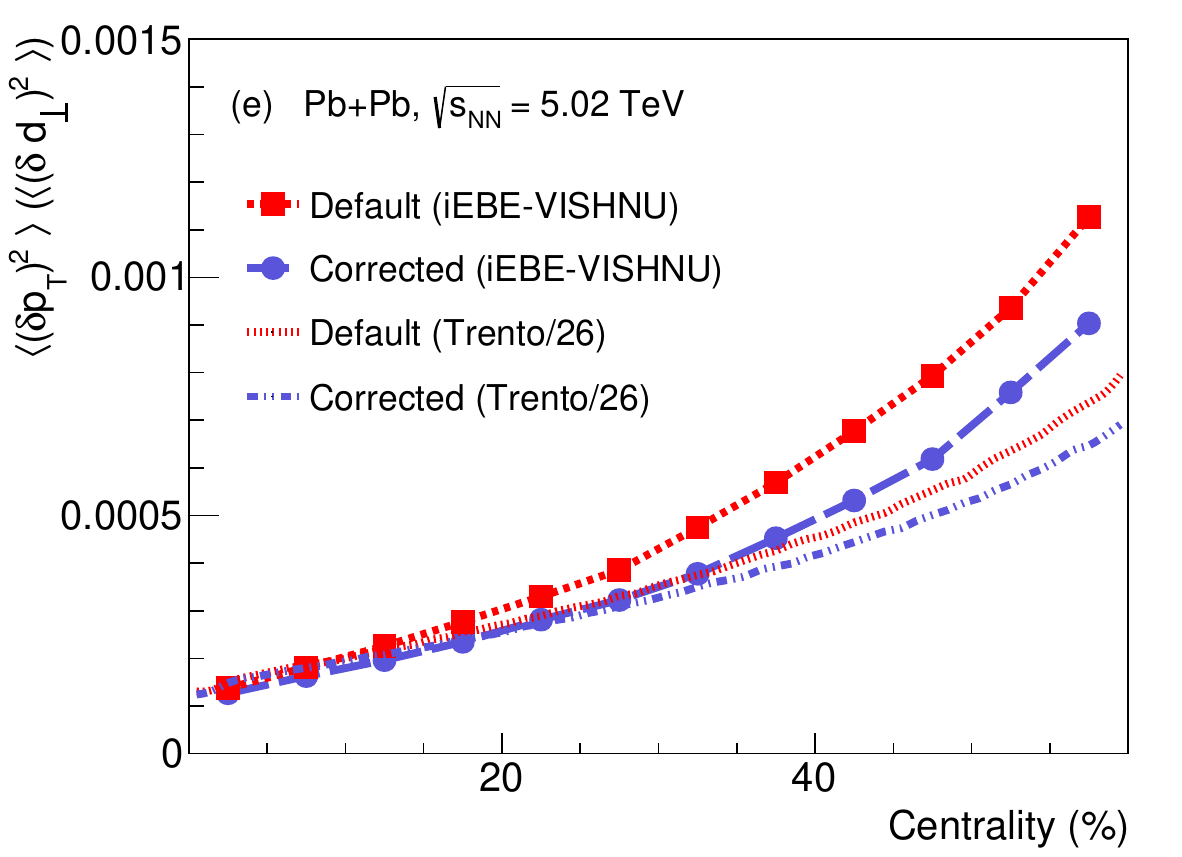}
\includegraphics[width=0.45\textwidth]{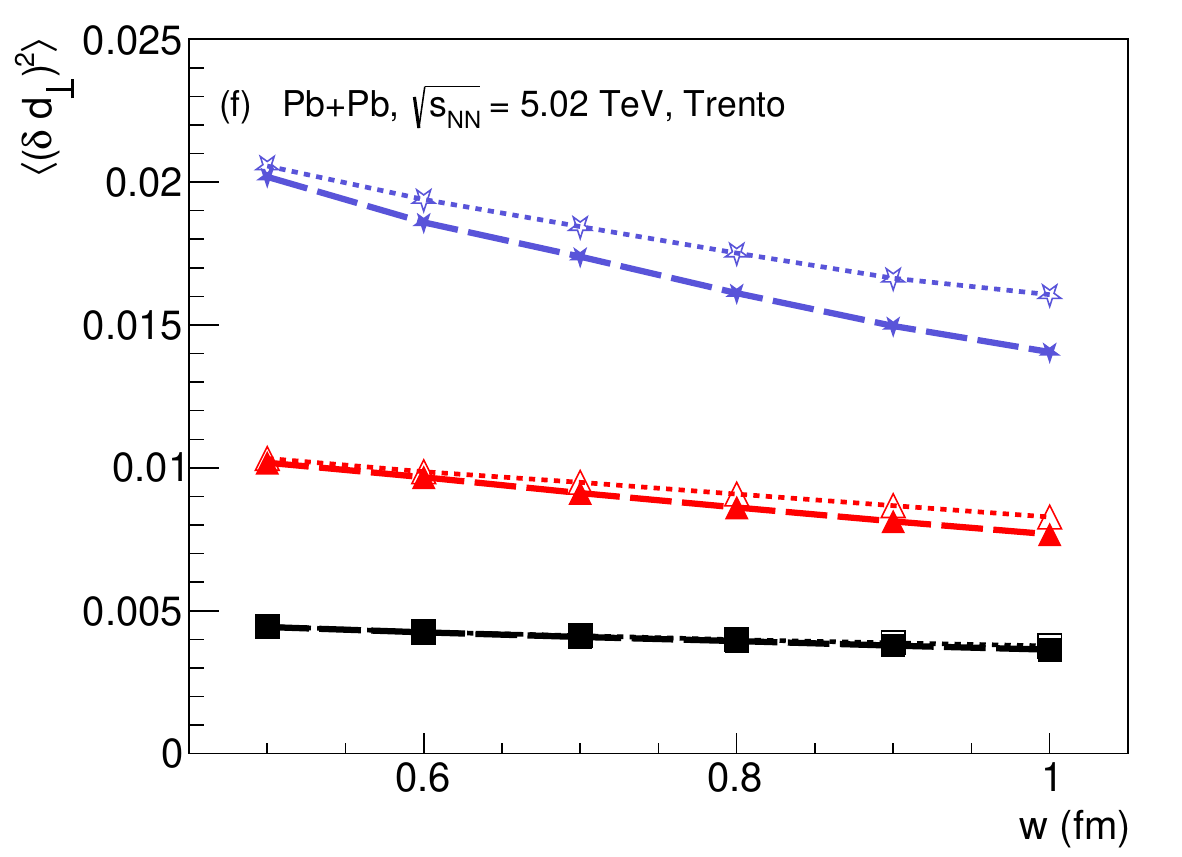}
\caption{(Color online) (Left panels) Centrality dependence of $v_{2}^{2}$, $v_{3}^{2}$ and $\deltapT$ with and without geometric inflation corrections in \PbPb\ collisions at $\snn=5.02 \ \mathrm{TeV}$ from \iebe\ simulations. The corresponding initial predictors from \trento\ are rescaled by a constant to match the magnitude of the final-state observables for shape comparison, as the hydrodynamic response is approximately linear. (Right panels) Nucleon-size dependence of the initial predictors $\varepsilon_{2}^{2}$, $\varepsilon_{3}^{2}$, and $\mean{(\delta d_\perp)^2}$ from \trento\ simulations.} \label{fig:flow}
\end{figure*}

\begin{figure*}[bht]
\centering
\includegraphics[width=0.45\textwidth]{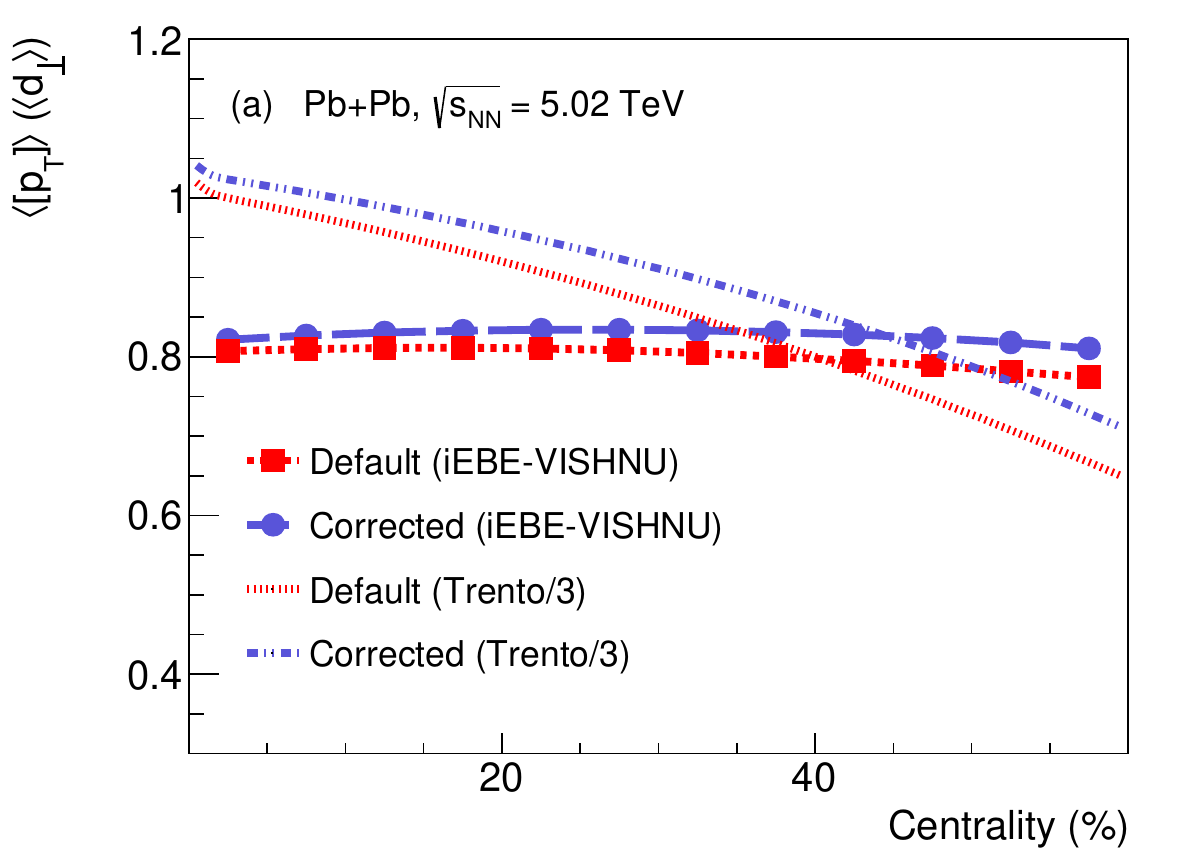}
\includegraphics[width=0.45\textwidth]{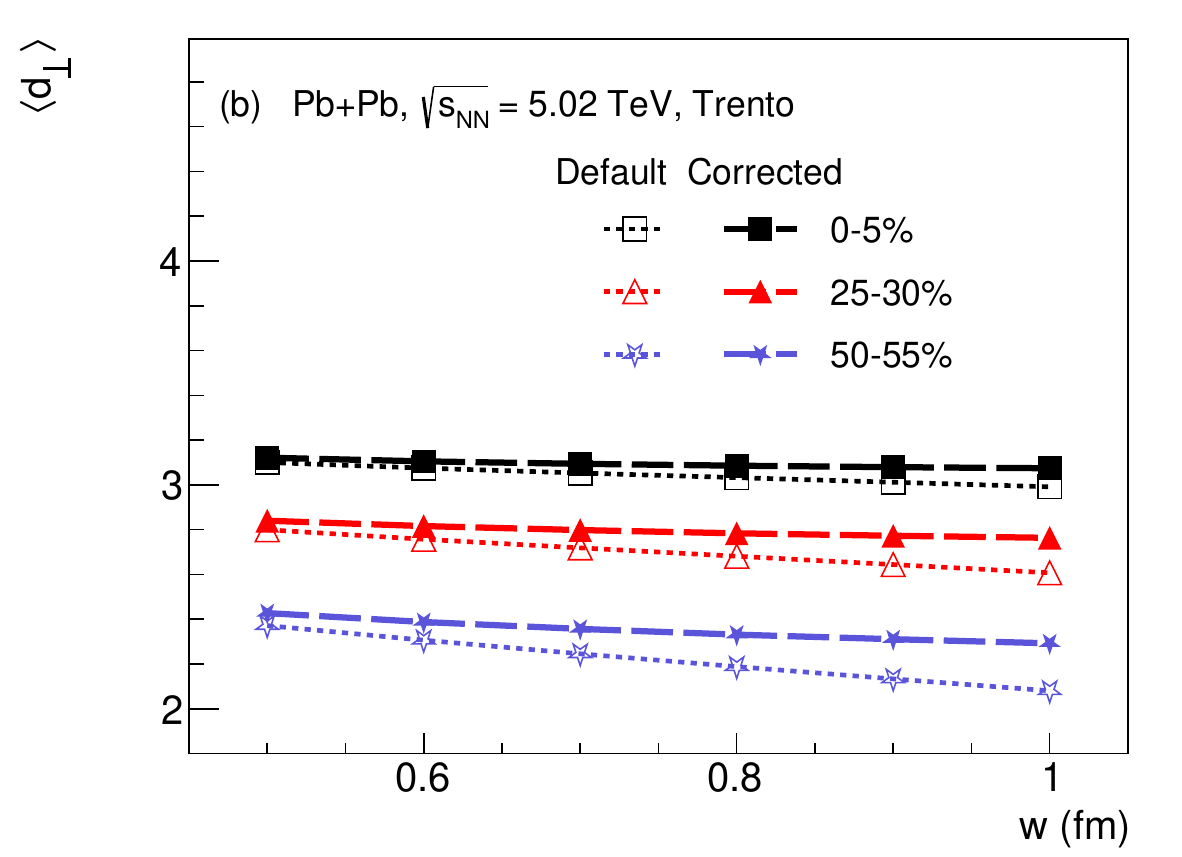}
\caption{(Color online) (a) Centrality dependence of $\meanpt$ with and without geometric inflation corrections in \PbPb\ collisions at $\snn=5.02 \ \mathrm{TeV}$ from \iebe\ simulations. The corresponding initial predictors are rescaled by a constant for shape comparison. (b) Nucleon-size dependence of the initial predictor $\mean{d_{\perp}}$ from \trento\ simulations.} \label{fig:meanpt}
\end{figure*}

\section{Results and Discussion}
\label{sec:discussion}

We first present results from full hydrodynamic simulations. Using the Bayesian-calibrated parameter set, the hydrodynamic model underestimates $\rhotwo$ in peripheral collisions with the traditional treatment~\cite{Giacalone:2021clp,ALICE:2021gxt,ATLAS:2019pvn}. To correct for the geometric inflation caused by the large nucleon width $w=0.92$ fm, the distribution of nucleon centers must have a significantly reduced diffuseness ($\tilde{a}=0.24$ fm) to compensate for the subsequent smearing. As shown in Fig.~\ref{fig:rhohydro}(a), this small $\tilde{a}$ value further reduces the predicted $\rhotwo$, widening the disagreement with experimental data in peripheral collisions. This suggests that either the true nucleon width is considerably smaller than Bayesian estimates, or that other missing physics (e.g., sub-nucleonic fluctuations) compensates. Conversely, the correction increases $\rhothree$, bringing the predictions into better agreement with data (Fig.~\ref{fig:rhohydro}c). This opposing behavior in two related observables indicates that while a simple rescaling of $w$ cannot simultaneously describe both, the self-consistent geometry provides a more physically motivated baseline for future refinements.

The opposing shifts in $\rhotwo$ and $\rhothree$ under the geometric inflation correction highlight a complex interplay between global nuclear geometry and event-by-event fluctuations. It is important to emphasize that the reduced diffuseness $\tilde{a}=0.24$ fm in the corrected sampling is \textit{not} a physical property of the nucleus; rather, it is a mathematical constraint required to recover the physical nuclear density after convoluting with a wide nucleon profile.

If we instead required the point-nucleon distribution to match the point-proton distribution inferred from the charge distribution ($R_{\rm point}=6.682$ fm, $a=0.447$ fm~\cite{Loizides:2017ack, Klos:2007is}), the inflated mass density with $w=0.92$ fm would imply a mass radius significantly larger than the charge radius:
\begin{equation}
R_{m}^{2} = R_{\rm ch}^{2} - r_{\rm ch,\,p}^{2} + 3w^2 \approx R_{\rm ch}^{2} + 1.8\, {\rm fm}^{2},
\end{equation}
where $r_{\rm ch,p}$ is the proton charge radius. Using the traditional parameters ($R=6.62$ fm, $a=0.546$ fm) for nucleon centers in \trento\ further exacerbates these differences. While recent Bayesian analyses treating proton and neutron densities separately account for some inflation effects~\cite{Nijs:2022rme}, the variation in $w$ still introduces additional inflation that significantly alters conclusions regarding hadronic observables~\cite{Xu:2026vnl}.

Traditionally, nucleon centers are held fixed while varying $w$, causing the mass density to change via geometric inflation. Our corrected method instead fixes the mass density, effectively decoupling it from QGP studies. For detailed investigations of nuclear structure (e.g., neutron skin measurements~\cite{Giacalone:2023cet}), a more realistic approach would use a consistent combination of nucleon center distributions and nucleon profile distributions derived from nuclear structure theory. We leave a full analysis for future work; here, we focus on the impact of the correction on flow observables.

For the key observable $\rhopearson$, we use the corresponding initial-state correlations $\rho_{n;\varepsilon}$ as predictors, defined as:
\begin{equation}
\rho_{n;\varepsilon} = \frac{\mathrm{cov}(\varepsilon_n\{2\}^2,d_\perp)}{\sqrt{\mathrm{var}(\varepsilon_n\{2\}^2)\,\mathrm{var}(d_\perp)}}.
\end{equation}
This mirrors the form of $\rhopearson$, with elliptic flow $v_2$ and $\pT$ replaced by the eccentricity $\varepsilon_n=-\{r^n e^{in\phi}\}/\{r^n\}$ and the initial energy per particle $d_\perp \equiv E/S$~\cite{Giacalone:2020dln}, based on the hydrodynamic linear-response relation~\cite{Schenke:2020uqq,Jia:2021qyu, Dimri:2023wup}. Here, $\delta d_\perp$ denotes fluctuations of the initial gradient $d_\perp$, $S=\int dxdy s(x,y)$ is the total multiplicity (entropy), and $E\approx \int dxdy s(x,y)^{4/3}$ is the total energy. The \trento\ model provides the initial entropy density $s(x,y)$ in the transverse plane. As shown in the left panels of Fig.~\ref{fig:rhohydro}, the predictors capture the relative differences between the two scenarios well, despite the sizable magnitude discrepancies. We therefore rely on these predictors to discuss the $w$-dependence between the two scenarios.

The \trento\ model simulation results for most-central to peripheral collisions are shown in the right panels of Fig.~\ref{fig:rhohydro}. The pronounced change in $\rho_{2}$ at peripheral collisions indicates that $\rho_{2}$ is sensitive to fluctuations of nucleon centers rather than the total mass density. For the most central collisions, the mass-density correction renders these observables largely insensitive to $w$, providing a robust probe of nuclear deformations~\cite{Bally:2021qys,Zhao:2024lpc,Giacalone:2020awm}. However, for $\rho_{3}$, the correlation with mass density makes it consistently less sensitive to the nucleon width $w$, especially at larger $w$. This contradicts the simple expectation that $\rho_{2}$ would be more sensitive to global geometry (total mass density) than $\rho_{3}$.

These findings motivate a detailed study of observables related to $\rhopearson$. Figure~\ref{fig:flow} shows the flow anisotropy coefficients $v_{n}^{2}$ and transverse momentum variance $\deltapT$ from \iebe\ simulations, alongside the $w$-dependence of initial predictors $\varepsilon_{n}^{2}$ and $\mean{(\delta d_{\perp})^{2}}$ from \trento\ simulations. Since flow observables were used to calibrate the \iebe\ model parameters, we do not overlay experimental data in these plots to avoid implying preference for the default scenario. Rather, our results call for a revision of Bayesian analyses under the inflation-corrected scenario.

\begin{figure*}[thb]
\centering
\includegraphics[width=0.45\textwidth]{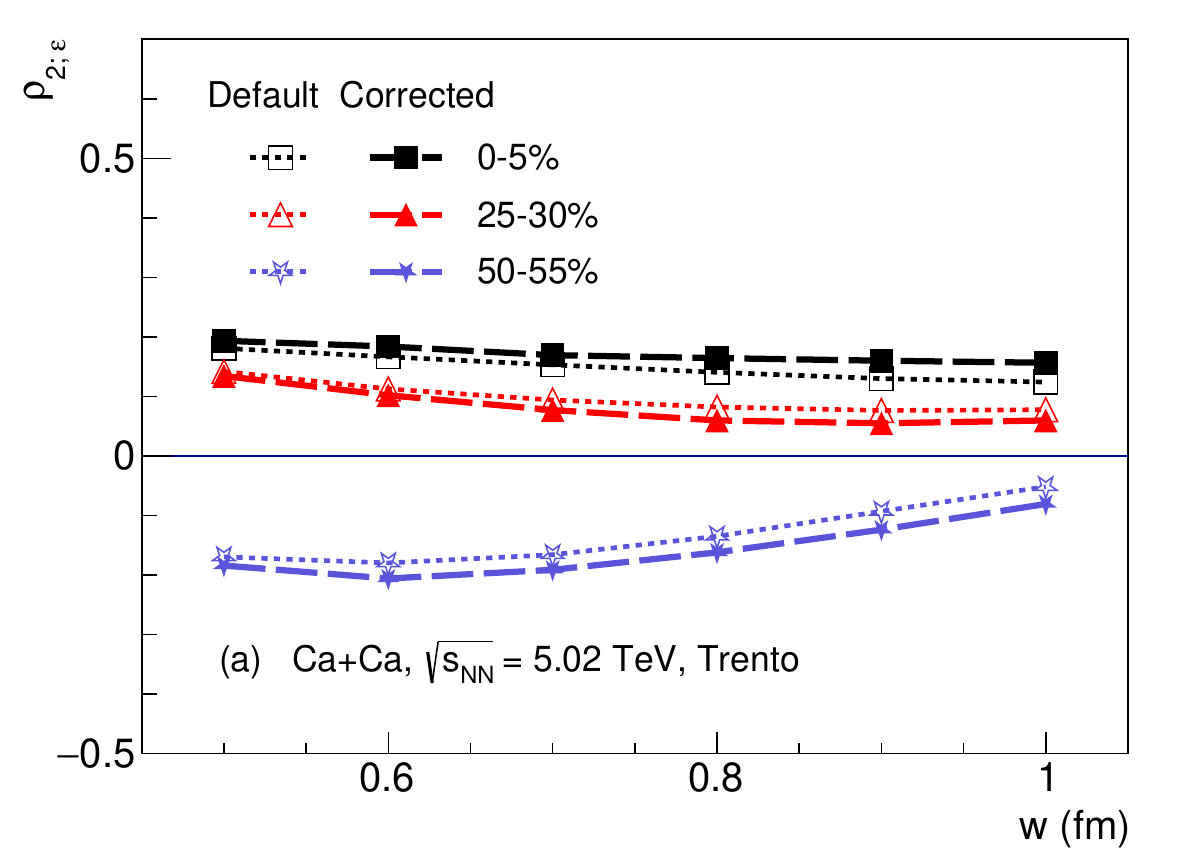}
\includegraphics[width=0.45\textwidth]{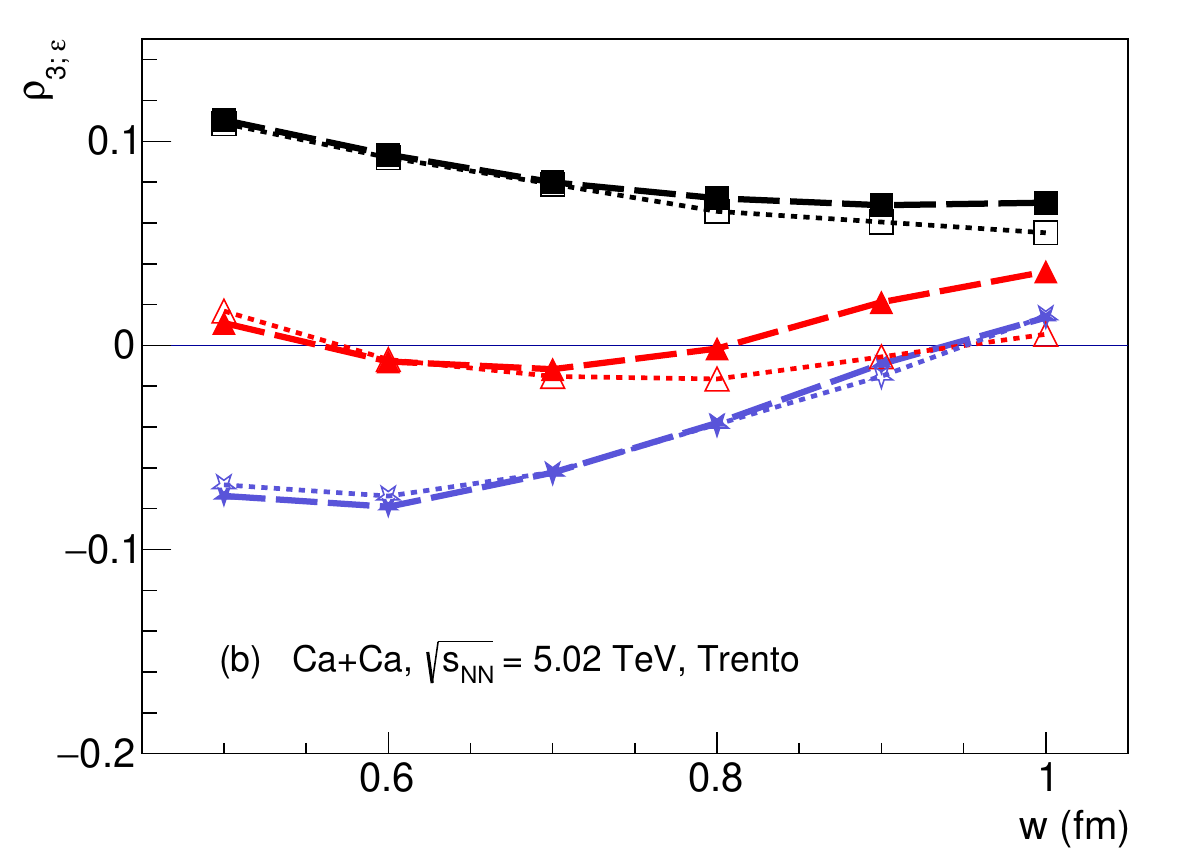}
\caption{(Color online) Nucleon-size dependence of initial predictors (a) $\rho_{2;\,\varepsilon}$ and (b) $\rho_{3;\,\varepsilon}$ in \CaCa\ collisions at $\snn=5.02$ TeV from \trento\ simulations.} \label{fig:cuvnpt}
\end{figure*}

\begin{figure*}[t]
\centering
\includegraphics[width=0.45\textwidth]{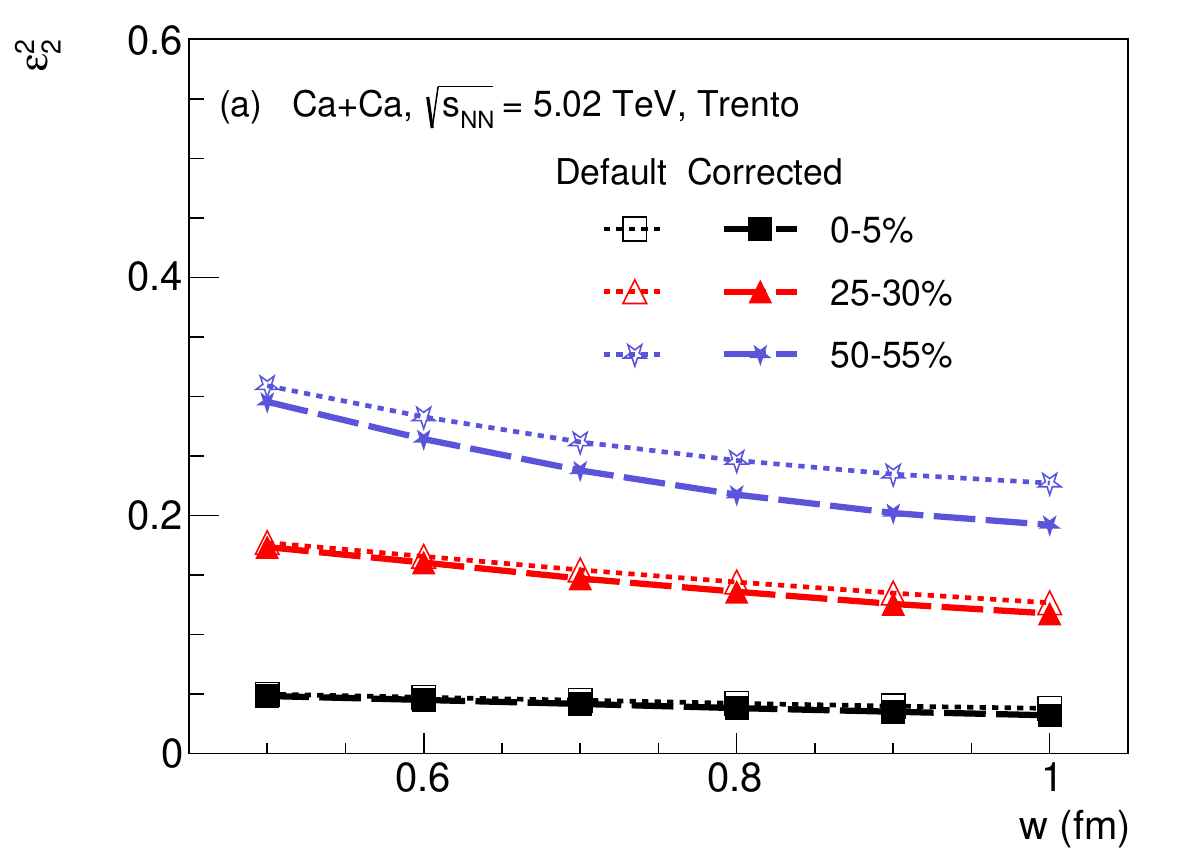}
\includegraphics[width=0.45\textwidth]{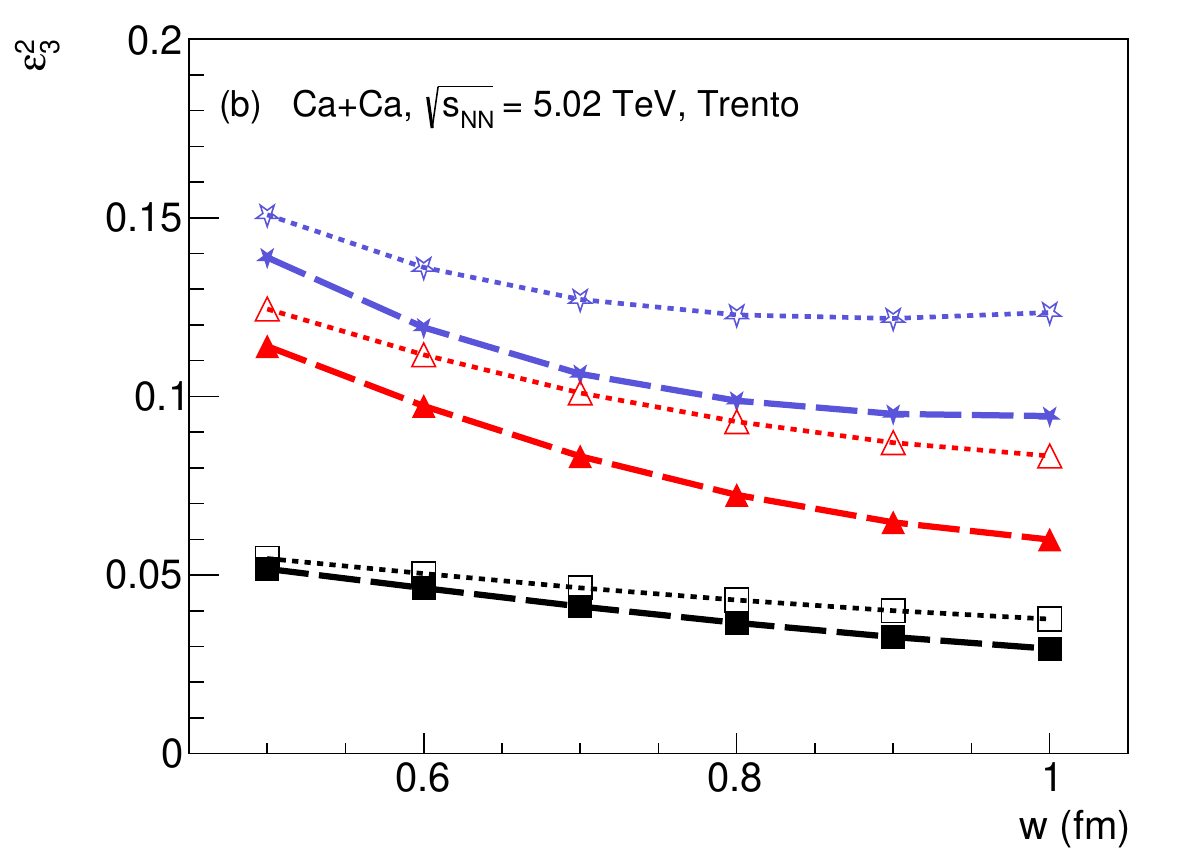}
\includegraphics[width=0.45\textwidth]{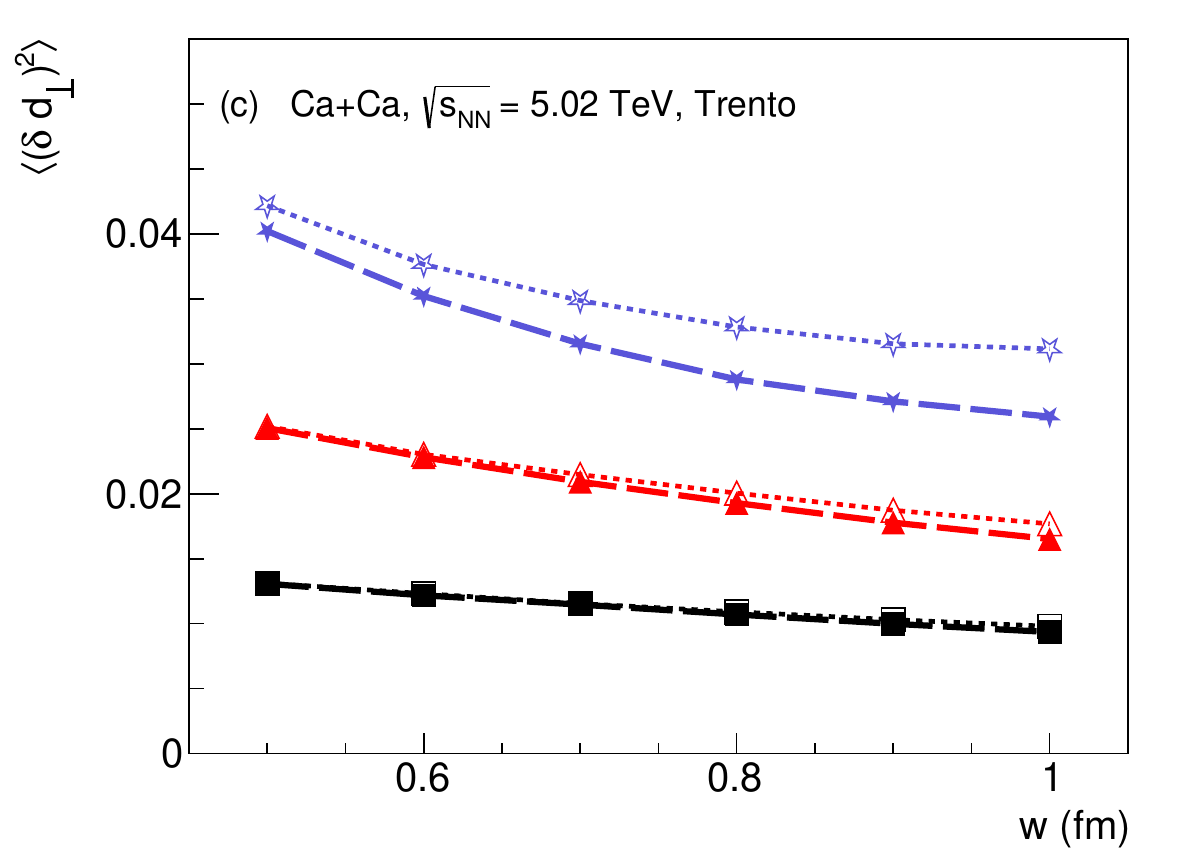}
\includegraphics[width=0.45\textwidth]{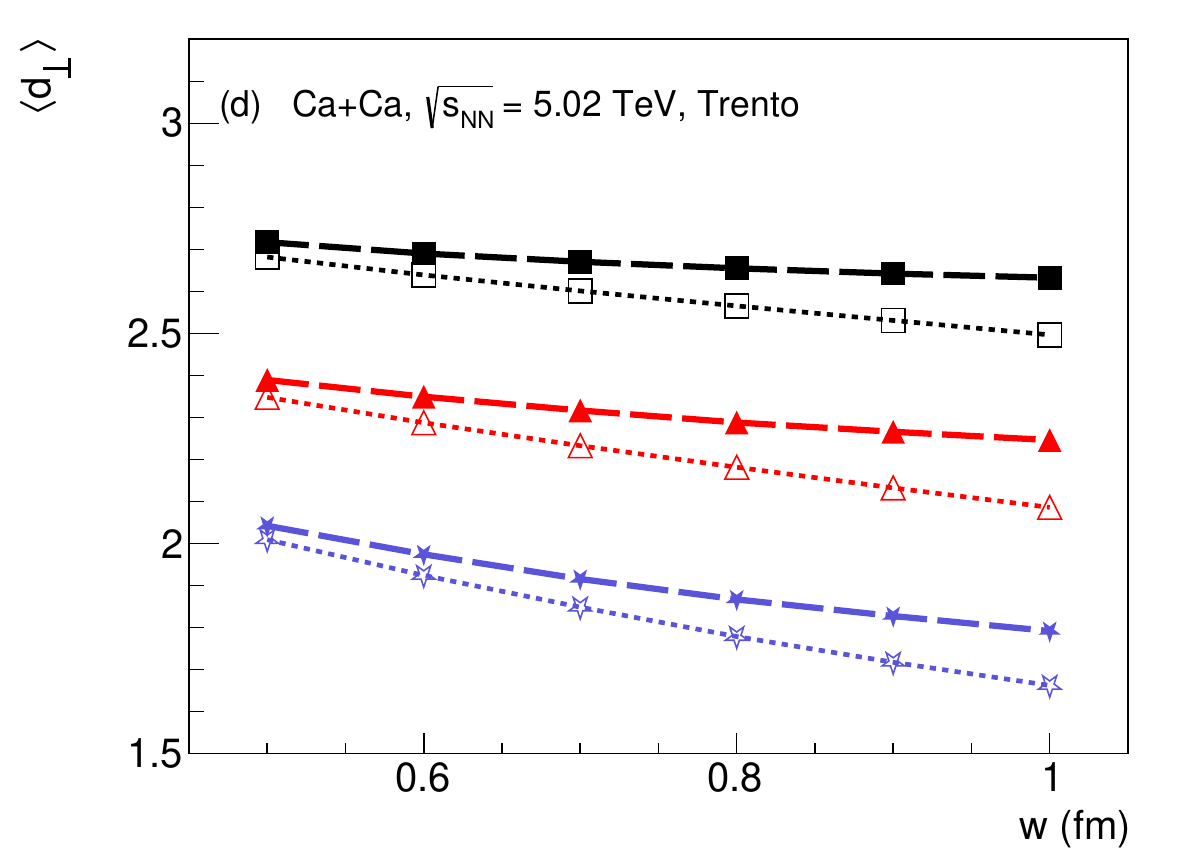}
\caption{(Color online) Nucleon-size dependence of initial predictors for \CaCa\ collisions at $\snn=5.02$ TeV: (a) elliptic eccentricity $\varepsilon_{2}^{2}$, (b) triangular eccentricity $\varepsilon_{3}^{2}$, (c) relative fluctuations of the initial gradient $\mean{(\delta d_\perp)^2}$, and (d) mean initial gradient $\mean{d_{\perp}}$.} \label{fig:cuflow}
\end{figure*}

As shown in the right panels of Fig.~\ref{fig:flow}, inflation corrections reduce the sensitivity of $v_{2}$ to the nucleon size parameter $w$, suggesting that elliptic flow primarily depends on the total mass density rather than on fluctuations of nucleon centers. This effect is especially pronounced at large $w$ in peripheral collisions. Conversely, for fluctuation-dominated observables like $v_{3}$ and $\deltapT$, inflation corrections enhance their sensitivity to nucleon size, as modifying the distribution of nucleon centers strongly affects their fluctuations. Their behavior confirms that these observables are sensitive to both nuclear geometry and its fluctuations, albeit to varying degrees. Differences in the diffuseness parameter $a$ have been utilized to study nuclear structure in relativistic isobar collisions~\cite{Li:2019kkh,Xu:2021vpn,Wang:2023yis,Jia:2022ozr}. 

The $\meanpt$ is known to be sensitive to the overall initial geometry and the resulting energy density~\cite{Broniowski:2009fm,Bozek:2012fw,Xu:2021uar}. As shown in Fig.~\ref{fig:meanpt}, geometric inflation systematically reduces the magnitude of $\meanpt$; because inflation increases the effective interaction area, it lowers the initial energy density per unit area. The inflation correction significantly reduces this unintentional sensitivity to the nucleon size parameter $w$, though a complete quantitative assessment requires further full hydrodynamic simulations. This observable remains a key candidate for constraining nuclear size and the neutron skin thickness in relativistic isobar collisions~\cite{Xu:2021uar}. While inflation effects can be partially canceled by taking ratios (e.g., \RuRu/\ZrZr), a precise determination of the neutron skin still necessitates a consistent treatment of the input nuclear densities.

Because the correction ties elliptic flow and $\meanpt$ more tightly to the fixed global density and enhances the sensitivity of $v_3$ and $\deltapT$ to nucleon-center fluctuations, we anticipate that a Bayesian re-analysis with the corrected geometry will alter the inferred posterior distributions of both initial-state and medium properties. In particular, the reduced response of $v_2$ and $\meanpt$ to $w$ may weaken the previous preference for large $w$, while the enhanced sensitivity of $v_3$ and $\deltapT$ could provide stronger constraints on the fluctuation scale. Re-evaluating calibrations such as JETSCAPE or Trajectum with inflation-corrected \trento\ profiles would be an essential next step to determine whether the ``large-$w$'' solution is genuinely an artifact of geometric inflation.

We investigate these fluctuation effects in \CaCa\ collisions at $\snn=5.02$ TeV, focusing on the $w$ dependence of the observables using \trento\ simulations. The default Woods-Saxon parameters for \Ca\ are $R=$ 3.723 fm and $a=$ 0.523 fm~\cite{Fricke:1995zz}.
After the inflation correction, the nucleon-center sampling parameters are listed in Tab.~\ref{tab:paras}, following the same procedure as for \Pb. In general, the impact on the Pearson coefficient mirrors that in \PbPb\ collisions, whereas the effect on nucleon-center fluctuations is suppressed, as shown in Fig.~\ref{fig:cuvnpt}.

For the elliptic eccentricity shown in Fig.~\ref{fig:cuflow}(a), the response to variations in $w$ and $a$ differs from the behavior observed in \PbPb\ collisions. This likely indicates that the dominant contribution arises from the small mass number of \Ca, which creates a high-fluctuation baseline that masks subtle geometric changes.

In contrast, for triangular eccentricity, mean transverse momentum, and its fluctuations (Fig.~\ref{fig:cuflow}(b--d)), the results are highly sensitive to the method of modification. Adjusting the total geometry via $w$ versus isolating the fluctuations via $a$ yields distinctly different outcomes, a trend analogous to that observed in large systems. Specifically, Fig.~\ref{fig:cuflow}(b) and (c) demonstrate that while $\varepsilon_3^2$ and $\mean{(\delta d_{\perp})^{2}}$ are driven by fluctuations, the ``inflation'' of the total nuclear volume significantly dampens the fluctuation signal. Figure~\ref{fig:cuflow}(d) further shows that the bulk predictor $\mean{d_\perp}$ is heavily influenced by whether the global density profile is preserved. This confirms that a self-consistent treatment of nuclear density is not a specialized requirement for large nuclei, but a crucial prerequisite for reliable modeling across all system sizes.

If an observable is sensitive to fluctuations rather than the overall geometry of the colliding nuclei, the sampling method governing these fluctuations must be chosen with care. Current geometry models assume identical nucleon profiles from the nuclear core to the surface. Under this approximation, the distribution of proton centers can be constrained directly from measured charge distributions. However, this fluctuation sampling can be significantly biased if nucleon profiles depend on local nuclear density. Such effects are expected to be especially prominent in light-nucleus collisions.

While this work establishes the qualitative importance of the geometric inflation correction, several extensions are necessary for quantitative precision. First, we have performed full hydrodynamic simulations only for a single large width $w=0.92$ fm; a systematic scan over $w$ with the corrected method would quantify the residual sensitivity and enable a rigorous Bayesian update. Second, the correction procedure has been applied only to spherical Woods-Saxon nuclei; extending it to deformed species (e.g., $^{238}$U) and to nuclei with experimentally measured charge densities would broaden its applicability~\cite{STAR:2024wgy,Giacalone:2020awm,Li:2026igf}. Third, we have assumed Gaussian nucleon profiles throughout; other functional forms (e.g., exponential, hard-sphere) may alter the required correction~\cite{Xu:2026vnl,Wang:1991hta,dEnterria:2020dwq}. These extensions represent important avenues for future work.

\section{Summary}
\label{sec:summary}

This study addressed the systematic artifact of ``geometric inflation'' in the initial-state modeling of relativistic heavy-ion collisions. The use of a finite nucleon size (width~$w$) in initial geometry models, via the convolution of point-like positions with a Gaussian profile, unintentionally inflates the nuclear density profile relative to the intended physical distribution. This artifact biases the initial geometry, subsequently affecting the calculated anisotropic flow harmonics~$\vnt$ and the event-by-event Pearson correlation~$\rhopearson$ between flow and mean transverse momentum fluctuations.

To isolate genuine nucleon-size effects from this unintentional rescaling, we implemented a self-consistent framework correcting for geometric inflation. By adjusting the sampling distribution of nucleon centers, we ensured that the global nuclear geometry remained invariant as~$w$ was varied. We focused on a representative large width~$w = 0.92$~fm, motivated by recent Bayesian analyses. Using the \iebe\ hybrid model for \PbPb\ collisions at $\snn = 5.02$~TeV, we compared observables calculated with standard (uncorrected) and corrected initial-state geometries.

Our key findings reveal that correcting for inflation fundamentally alters the sensitivity of observables. Elliptic flow~$v_2$ and mean transverse momentum $\meanpt$ become more strongly tied to the fixed global nuclear geometry (mass density), rendering them less sensitive to variations in~$w$. Conversely, triangular flow~$v_3$  and $\deltapT$ exhibit heightened sensitivity to event-by-event fluctuations of nucleon positions. The correlation coefficient~$\rhotwo$ is significantly reduced in peripheral collisions under the corrected geometry, whereas~$\rhothree$ increases. This divergent behavior underscores the complex interplay between global geometry and fluctuations. Crucially, while the strong sensitivity of~$\rhopearson$ to~$w$ persists, the baseline is shifted by the correction; thus, inferring the nucleon width~$w$ without accounting for geometric inflation risks biased results. Comparative studies with smaller \CaCa\ systems show that while the effects on Pearson coefficients are similar, finite-number fluctuations may play a more dominant role.

In conclusion, geometric inflation is a non-negligible artifact in initial-state models incorporating finite nucleon size. Its correction effectively decouples the description of overall nuclear geometry from the granularity introduced by nucleon width, leading to a qualitatively distinct interpretation of flow observables. This work establishes that a self-consistent treatment of the initial nuclear density is essential for reliably using heavy-ion collision data to constrain fundamental nucleon properties and for performing unbiased Bayesian extraction of quark-gluon plasma transport coefficients. Future analyses should adopt such corrected frameworks to break degeneracies between initial-state and medium parameters.

\section{Acknowledgments}
This work is supported in part by the National Natural Science Foundation of China under Grants No.~12275082.

\bibliography{ref}

\end{document}